\documentclass[useAMS,usenatbib,usegraphicx]{mn2e}

\title[Optical pulsations from AXP 1E\,1048.1--5937]{Optical pulsations from the anomalous X-ray pulsar 1E\,1048.1--5937}

\author[V. S. Dhillon et al.]{V. S. Dhillon,$^1$\thanks{E-mail: vik.dhillon@sheffield.ac.uk}
T. R. Marsh,$^2$ S. P. Littlefair,$^1$ C. M. Copperwheat,$^2$ P. Kerry,$^1$
\newauthor
R. Dib,$^3$ M. Durant,$^4$ V. M. Kaspi,$^3$ R. P. Mignani,$^5$ A. Shearer$^6$ \\\\
$^{1}$Department of Physics and Astronomy, University of Sheffield, 
Sheffield S3 7RH, UK \\
$^{2}$Department of Physics, University of Warwick, Coventry CV4 7AL, UK \\
$^{3}$Department of Physics, McGill University, Montreal, Quebec H3A 2T8, Canada \\
$^{4}$Instituto de Astrof\'\i{}sica de Canarias, 38200 La Laguna, Tenerife, Spain \\
$^{5}$Mullard Space Science Laboratory, University College London, Holmbury St. Mary, Dorking, Surrey, RH5 6NT, UK \\
$^{6}$Centre for Astronomy, National University of Ireland, Galway, Newcastle Rd., Galway, Ireland \\}
\begin{document}

\date{Submitted on 2008 November 12.}

\pagerange{\pageref{firstpage}--\pageref{lastpage}} \pubyear{2008}

\maketitle

\label{firstpage}

\begin{abstract}
  We present high-speed optical photometry of the anomalous X-ray
  pulsar 1E\,1048.1--5937 obtained with ULTRACAM on the 8.2-m Very
  Large Telescope in June 2007. We detect 1E\,1048.1--5937 at a
  magnitude of $i'=25.3\pm0.2$, consistent with the values found by
  \citet{wang08} and hence confirming their conclusion that the source
  was approximately 1 mag brighter than in 2003--2006 due to an
  on-going X-ray flare that started in March 2007. The increased
  source brightness enabled us to detect optical pulsations with an
  identical period (6.458\,s) to the X-ray pulsations. The rms pulsed
  fraction in our data is $21\pm7$\%, approximately the same as the
  2--10 keV X-ray rms pulsed fraction. The optical and X-ray pulse
  profiles show similar morphologies and appear to be approximately in
  phase with each other, the latter lagging the former by only
  $0.06\pm0.02$ cycles. The optical pulsations in 1E\,1048.1--5937 are
  very similar in nature to those observed in 4U\,0142+61. The
  implications of our observations for models of anomalous X-ray
  pulsars are discussed.
\end{abstract}


\section{Introduction}

The anomalous X-ray pulsars (AXPs) are a small group\footnote{See {\tt
http://www.physics.mcgill.ca/$\sim$pulsar/magnetar/main.html} for an
up-to-date catalogue of all known AXPs, including the various
wavelengths at which each has been detected.} of isolated neutron
stars in which the X-ray luminosity far exceeds the energy available
from the spin-down. The AXPs are generally believed to be magnetars,
in which the excess luminosity is powered by the decay of an
ultra-strong magnetic field, in excess of $10^{14}$\,G (see
\citealt{woods06}). An alternative explanation is the fallback disc
scenario, in which some of the supernova ejecta fails to escape and
forms an accretion disc around the neutron star, providing an extra
source of energy to power the X-ray emission (\citealt{vanparadijs95};
\citealt{chatterjee00}; \citealt{alpar01}).

One way of discriminating between the magnetar and fallback disc
models is via optical observations. The magnetar model predicts any
optical emission must be non-thermal and magnetospheric in origin.
Four plausible mechanisms have been considered -- coherent plasma
emission, synchrotron emission from electrons with high Lorentz
factors, cyclotron emission from ions in the outer magnetosphere and
curvature emission from bunched electron-positron pairs in the inner
magnetosphere (see \cite{beloborodov07} and references therein). The
fallback disc model, on the other hand, predicts any optical emission
is produced by reprocessing of the X-ray light in the disc and/or
thermal emission from the disc \citep{perna00}.

The first AXP to be detected in the optical part of the spectrum was
4U\,0142+61 \citep{hulleman00}. Optical pulsations were discovered in
4U\,0142+61 by \cite{kern02b}, and the fact that these pulsations have
the same period, morphology and phase as the X-rays, but with 5--7
times greater pulsed fraction, was reported by \cite{dhillon05}. These
results provided strong support for the magnetar model -- pulsed
optical emission is indicative of a magnetospheric origin and disc
reprocessing is unlikely in this case, as the optical pulsed fraction
is higher than the X-ray pulsed fraction and there is no time delay
between the two. Although it is possible to contrive ways in which the
fallback disc model is consistent with the optical pulsations observed
in 4U\,0142+61, e.g. by assuming that the X-ray pulse profile that we
observe is different to the X-ray radiation seen by the disc due to
orientation or beaming effects, or by invoking a hybrid
disc-magnetosphere model (see \cite{ertan07} and references therein),
the weight of evidence from the optical and other wavelengths lies
heavily on the side of the magnetar model (see \cite{mereghetti08} for
a recent review).

The detection of optical pulsations in other AXPs would provide
valuable confirmation, or otherwise, of the results for 4U\,0142+61
discussed above. Only one other AXP has been unambiguously identified
in the optical: 1E\,1048.1--5937 \citep{durant05}$^1$. In this paper
we report on the first detection of optical pulsations from AXP
1E\,1048.1--5937, obtained only $\sim$\,3 months after a bright X-ray
flare in March 2007.

\section{Observations and data reduction}
\label{obsred}

\begin{figure}
\centering
\includegraphics[width=5.2cm,angle=270]{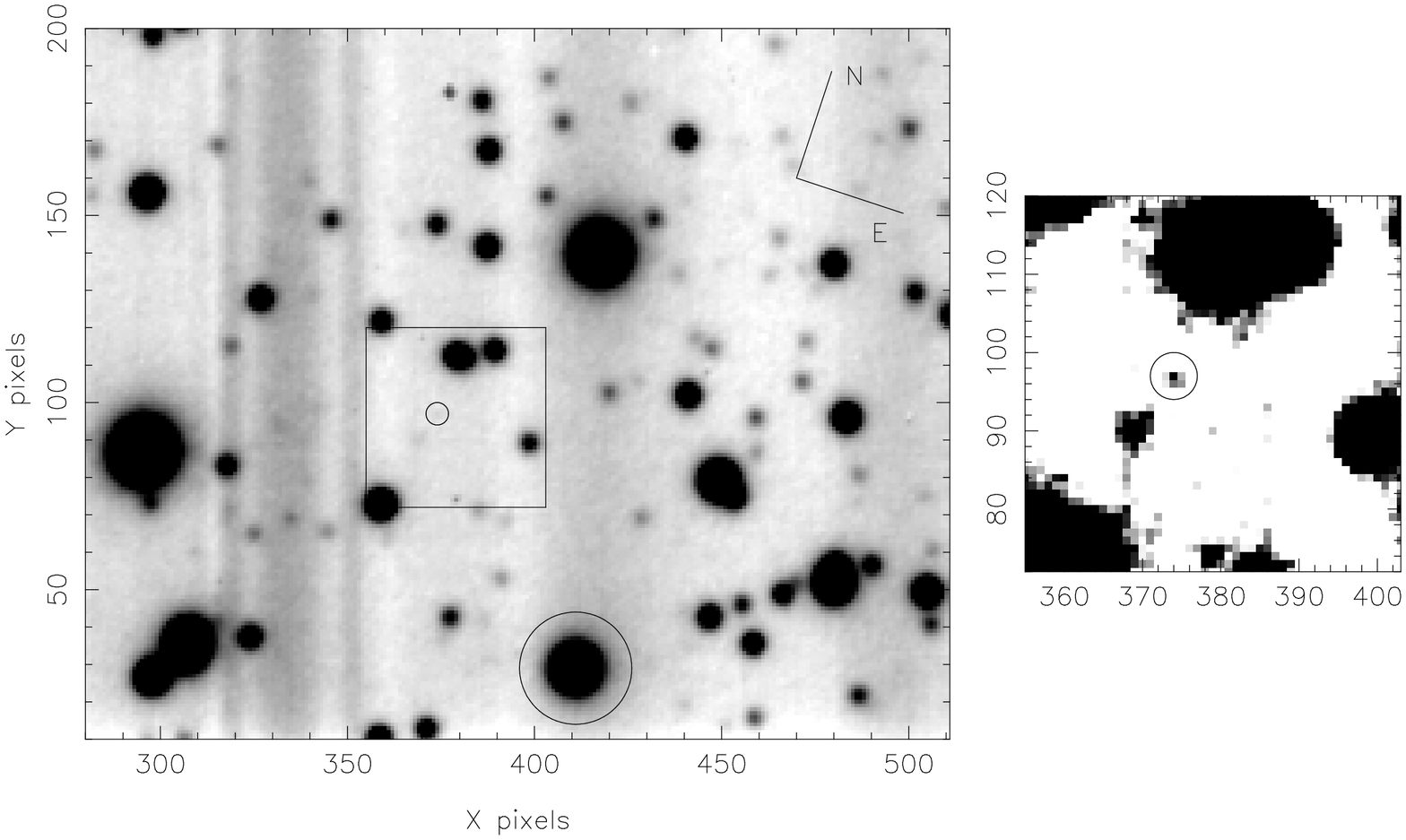}
\caption{Left: Summed $i'$-band image of the field around
  1E\,1048.1--5937, with a total exposure time of 10\,684~s.  For
  clarity, only a portion of one of the two ULTRACAM windows is shown.
  The positions of 1E\,1048.1--5937 and the comparison star are
  indicated by circles near the centre and bottom of the image,
  respectively. The central box shows the portion of the field that is
  plotted at a higher contrast on the right. The orientation of the
  image is marked on the upper right-hand side. The pixel scale is
  0.156 arcseconds/pixel, hence the field of view in this image is
  $36\times30$ arcseconds. The vertical banding is due to residual
  bias structure. Right: Higher contrast plot of a $7.5\times7.5$
  arcsecond field around 1E\,1048.1--5937, highlighting the detection
  of the pulsar in the $i'$-band}
\label{fig1}
\end{figure}

The observations of 1E\,1048.1--5937 presented in this paper were
obtained with ULTRACAM \citep{dhillon07b} at the Nasmyth focus of
Melipal, the 8.2-m Unit 3 of the Very Large Telescope (VLT) in
Chile. ULTRACAM is a CCD camera designed to provide imaging photometry
at high temporal resolution in three different colours
simultaneously. The instrument provides a 2.66 arcminute field on its
three $1024\times1024$ E2V 47-20 CCDs (i.e. 0.156
arcseconds/pixel). Incident light is first collimated and then split
into three different beams using a pair of dichroic beamsplitters. For
the observations presented here, one beam was dedicated to the SDSS
$u'$ ($\lambda_{eff}=3543$\AA) filter, another to the SDSS $g'$
(4770\AA) filter and the third to the SDSS (7625\AA) $i'$
filter. Because ULTRACAM employs frame-transfer chips, the dead-time
between exposures is negligible: we used ULTRACAM in its two-windowed
mode, each of $250\times200$ pixels, resulting in an exposure time of
0.963~s and a dead-time of 0.024~s. A total of 11\,095 frames of
1E\,1048.1--5937 were obtained on the night of 2007 June 9, with each
frame time-stamped to a relative (i.e. frame-to-frame) accuracy of
$\sim$\,50\,$\mu$s and to an absolute accuracy of $\sim$\,1 ms using a
dedicated GPS system (see \citealt{dhillon07b}). Observations of the
SDSS standard G163-51 \citep{smith02} were also obtained to flux
calibrate the data. The night was photometric, with no moon and
$i'$-band seeing of 0.65 arcseconds. The sum of the 11\,095 frames in
the $i'$-band is shown in figure~\ref{fig1}, which can be compared to
the finding charts presented by \cite{durant05}.

The data were reduced using the ULTRACAM pipeline software
\citep{dhillon07b}. All frames were first debiased and then
flat-fielded, the latter using the median of twilight sky frames taken
with the telescope spiralling. Adopting the same successful approach
that we used in our study of AXP 4U\,0142+61 \citep{dhillon05}, we
extracted light curves of 1E\,1048.1--5937 using two different
techniques:

\subsection{Technique (i)}
\label{tech1}

\begin{table}
\centering
\caption{
  X-ray ephemeris for 1E\,1048.1--5937 \citep{dib08}. The epoch of the
  frequency and frequency derivative measurements given below falls on the
  same night as our VLT observations (09/06/2007 = MJD 54\,260). BMJD refers 
  to the Barycentric-corrected Modified Julian Date on the Barycentric
  Dynamical Timescale (TDB). The errors on the last two digits of each
  parameter are given in parentheses. This ephemeris is valid for
  BMJD $54\,229.0-54\,280.0$.}
\begin{tabular}{lr}
& \\
\hline
\hspace*{4cm} & \\
$\nu$ (Hz)\dotfill & 0.1548479469(42) \\
$\dot\nu$ ($10^{-13}$ Hz s$^{-1}$)\dotfill & --5.413(53) \\
Epoch (BMJD)\dotfill & 54\,260.0 \\
& \\
\hline
\end{tabular}
\end{table}

As part of a long-term monitoring project, 1E\,1048.1--5937 has been
observed regularly (up to three times per week) since 1997 with the
Proportional Counter Array (PCA) on board the Rossi X-ray Timing
Explorer (RXTE) [\citealt{kaspi01}; \citealt{gavriil04};
\citealt{dib08}]. The X-ray spin frequency and frequency derivative of
1E\,1048.1--5937 for the night of our VLT observations (MJD 54\,260)
are given in table~1. For the first light-curve extraction technique,
we shifted and added each of the 11\,095 ULTRACAM frames into 10
evenly-spaced phase bins using the epoch and spin frequency given in
table~1, resulting in 10 high signal-to-noise data frames. An optimal
photometry algorithm \citep{naylor98} was then used to extract the
counts from 1E\,1048.1--5937 and an $i'$\,$\sim$\,17 comparison star
$\sim$\,12 arcseconds to the south-east of the AXP (see
figure~\ref{fig1}), the latter acting as the reference for the profile
fits and transparency-variation correction. The position of
1E\,1048.1--5937 relative to the comparison star was determined from a
sum of all the images, and this offset was then held fixed during the
reduction so as to avoid aperture centroiding problems. The sky level
was determined from a clipped mean of the counts in an annulus
surrounding the target stars and subtracted from the object counts.

\subsection{Technique (ii)}
\label{tech2}

The second approach we took to light curve extraction was identical to
that described above, except we omitted the phase-binning step and
simply performed optimal photometry on the 11\,095 individual ULTRACAM
data frames followed by a periodogram analysis of the resulting time
series. In other words, we made no assumption about the spin period of
1E\,1048.1--5937.

\section{Results}
\label{results}

\subsection{Magnitudes}
\label{mags}

We were unable to detect 1E\,1048.1--5937 in $u'$ and $g'$, at a
3$\sigma$ detection limit of $u'>25.7$ and $g'>27.6$, respectively.
This is unsurprising given the high visual extinction to the object
($A_V=4.9$; \citealt{durant06a}). We did, however, clearly detect
1E\,1048.1--5937 in $i'$ at a magnitude of $i'=25.3\pm0.2$, as shown
in figure~\ref{fig1}. \cite{durant05} found that 1E\,1048.1--5937 was
at a magnitude of $I=26.2\pm0.4$ ($i'$\,$\sim$\,26.1)\footnote{Noting that
the visual extinctions to 1E\,1048.1--5937 and 4U\,0142+61 are
approximately the same \citep{durant06a}, we have assumed that the
colours of 1E\,1048.1--5937 are the same as 4U\,0142+61
(\citealt{hulleman04}; \citealt{dhillon05}) and then used the
equations of \cite{smith02} to convert from $I$ to $i'$.} on
06/06/2003, just over one year after its first X-ray flare was
observed in April 2002.  \citet{wang08} observed 1E\,1048.1--5937 at a
magnitude of $I=24.9\pm0.2$ ($i'$\,$\sim$\,24.8)$^2$ on 07/05/2007, just over
one month after its second X-ray flare began on 2007 March 21. These
authors also measured a limit of $i'>24.5$ on 15/07/2007. Our
$i'$-band magnitude, which is slightly fainter and was obtained
slightly later than the first $I$-band measurement of \citet{wang08},
implies that we observed 1E\,1048.1--5937 whilst it was still
relatively bright, but declining, from the most recent X-ray
flare (see \citealt{tam08}).

\subsection{Pulse profiles}

The two data reduction techniques described in section~\ref{obsred}
result in two different pulse profiles for 1E\,1048.1--5937.

\subsubsection{Technique (i)}
\label{tech2ii}

\begin{figure}
\centering
\includegraphics[width=6cm,angle=270]{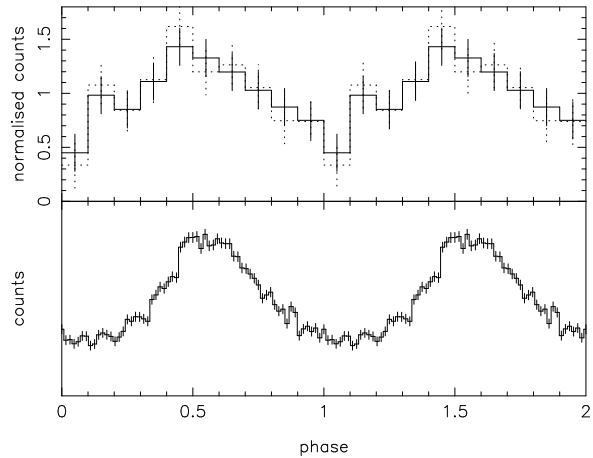}
\caption{Top: The solid and dotted lines show optical pulse profiles
  of 1E\,1048.1--5937 in the $i'$-band obtained using techniques (i)
  and (ii), respectively (see
  sections~\protect{\ref{tech1}}~and~\protect{\ref{tech2}}). Each
  pulse profile was first corrected for transparency variations using
  the comparison star shown in figure~\ref{fig1}, although the
  correction made only a negligible difference to the light
  curves. The pulse profiles were then normalised by dividing by the
  mean number of counts. For clarity, two cycles are shown. Bottom:
  Averaged X-ray pulse profile of 1E\,1048.1--5937 in the 2--10 keV
  energy band spanning the epoch of the optical observations
  (\citealt{dib08}). Note that it is not possible to estimate the
  X-ray pulsed fraction from this profile as the PCA on RXTE has a
  1$^{\circ}$ field of view and no imaging capability, rendering the
  background level uncertain. For this reason, no scale is given on
  the ordinate.}
\label{fig2}
\end{figure}

The first technique produced the pulse profile shown by the solid line
in the top panel of figure~\ref{fig2}. The pulse profile exhibits a
broad, single-humped structure with a peak around phase 0.5 and a
minimum around phase 0. There is a great deal of similarity in the
morphologies of the optical pulse profile shown in the top panel of
figure~\ref{fig2} and the 2--10 keV X-ray pulse profile shown below
it, where the latter is the average of the X-ray light curves in the
period 23/05/2007--21/06/2007 obtained as part of the RXTE monitoring
campaign described by \citet{dib08}, with a total effective
integration time of 59.5\,ks.  Both profiles show the same broad,
single-humped morphology. Moreover, since the X-ray light curve shown
in figure~\ref{fig2} has also been phased using the ephemeris given in
table~1, it can be seen that the optical and X-ray pulse profiles are
approximately in phase with each other. To quantify this, the optical
pulse profile was cross-correlated with the X-ray pulse profile. The
resulting peak in the cross-correlation function was fitted with a
Gaussian to derive a phase shift of $-0.06\pm0.02$ cycles
(i.e. $-0.39\pm0.13$\,s), where a negative phase shift implies that
the X-ray pulse profile lags the optical pulse profile. This phase
shift is only significant at the 3$\sigma$ level, due to the low
signal-to-noise and time resolution of the optical data, and
additional data will be required in order to confirm that the phase
shift is different from zero (discounting the unlikely situation in
which the time delay is approximately equal to some multiple of the
spin period).

It should be noted that the morphology of the X-ray pulse profile in
1E\,1048.1--5937 does not appear to be energy sensitive; the shapes of
the 2--4 keV and 6--10 keV pulse profiles are virtually identical to
the 2--10 keV pulse profile shown in figure~\ref{fig2}, even though
the 6--10 keV band is composed primarily of non-thermal photons
whereas the 2--4 keV band is composed of both thermal and non-thermal
photons (F. Gavriil, private communication).

The modulation amplitude of the pulses presented in figure~\ref{fig2}
can be measured using a peak-to-trough pulsed fraction, $h_{pt}$,
defined as follows:

\begin{equation}
h_{pt}=\frac{F_{max}-F_{min}}{F_{max}+F_{min}},
\end{equation}

\noindent where $F_{max}$ and $F_{min}$ are the maximum and minimum
flux in the pulse profile, respectively. We find a value of
$h_{pt}=52\pm15$\%. The peak-to-trough pulsed fraction defined in equation~1
effectively adds any noise present in the light curve to the true
pulsed fraction, thereby tending to increase the resulting
measurement. A more robust estimate is given by the root-mean-square
(rms) pulsed fraction, $h_{rms}$, defined as follows:

\begin{equation}
h_{rms}= \frac{1}{\bar{y}} \left[ \frac{1}{n}
\sum_{i=1}^{n}(y_i-\bar{y})^2-\sigma_i^2 \right]^{\!\frac{1}{2}},
\end{equation}

\noindent where $n$ is the number of phase bins per cycle, $y_i$ is
the number of counts in the $i^{\rm th}$ phase bin, $\sigma_i$ is the
error on $y_i$ and $\bar{y}$ is the mean number of counts in the
cycle. As expected, measuring the optical pulsed fraction in this way
gives a lower value of $h_{rms} = 21\pm7$\%.

For comparison, the 2--10 keV X-ray rms pulsed fraction at the same
epoch as the optical observations was $h_{rms} = 28.7\pm0.5$\%. Since
it isn't possible to measure the rms pulsed fraction directly from the
RXTE pulse profile due to the uncertain background (see caption to
figure~\ref{fig2}), we derived this value as follows.  We first
averaged the rms pulsed flux, defined as the product of the total flux
and the rms pulsed fraction, measured with RXTE between 2007 June 7
and June 12 \citep{dib08}. This value, which is background
independent, was identical to that measured with {\em Chandra} on 2007
April 28 by \cite{tam08}, who also find a strong anticorrelation
between total flux and rms pulsed fraction. Hence if the pulsed flux
was the same for the RXTE and {\em Chandra} observations, we can be
confident that the rms pulsed fraction was the same as well, and hence
we have adopted the {\em Chandra} rms pulsed fraction from the 2007
April 28 observation of \cite{tam08}.

\subsubsection{Technique (ii)}

\begin{figure}
\centering
\includegraphics[width=6cm,angle=270]{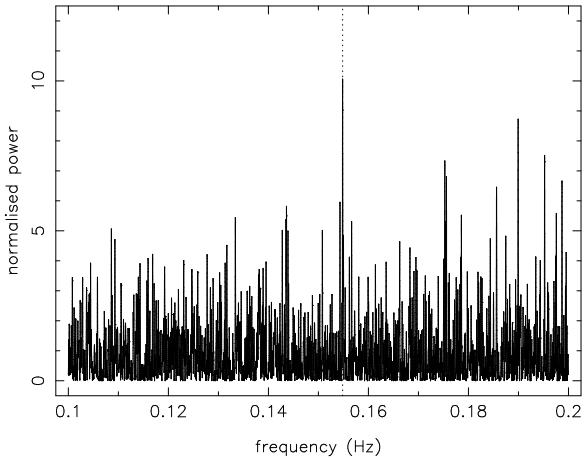}
\caption{Lomb-Scargle periodograms of 1E\,1048.1--5937 in the $i'$-band,
  obtained using the light curves from technique (ii)
  [section~\protect{\ref{tech2}}]. The dotted line shows the
  X-ray pulse frequency of 0.1548479469 Hz given in
  table~1.}
\label{fig3}
\end{figure}

The second data reduction technique (section~\ref{tech2}) can be used
to provide a check on the reliability of the optical pulse profile
derived using technique (i). Rather than adopting the X-ray ephemeris
given in table~1, we can instead determine the pulse period directly
from our optical data using a periodogram. Measuring the same period
in the optical and the X-rays would prove without doubt that we have
detected optical pulsations from 1E\,1048.1--5937.

Figure~\ref{fig3} shows the Lomb-Scargle periodograms \citep{press89}
for the 11\,095 points in the $i'$-band light curves. The light curve
was first corrected for transparency variations. The highest peak in
the resulting periodogram occurs at a frequency of
$0.15484\pm0.00004$~Hz ($6.458\pm0.002$~s), where the error is given
by the width ($\sigma$) of a Gaussian fit to the peak in the
periodogram. This frequency is consistent to the fifth decimal place
with the X-ray pulse frequency given in table~1, thereby confirming
that we have indeed detected the X-ray pulsation of 1E\,1048.1--5937
in the optical. We further tested the robustness of our period
detection by constructing 10\,000 randomised light curves from the
original light curves by randomly re-ordering the time-series.  None
of the resulting 10\,000 periodograms showed a higher peak at
0.15484~Hz. To check for artifacts, we calculated the Lomb-Scargle
periodograms of both the sky and the comparison star -- neither showed
any evidence for a periodicity at 0.15484~Hz, or at any other
frequency. We also searched for evidence of optical bursts and/or
longer-timescale periodicities in our 10\,684\,s light curve --
nothing significant was found.

Folding the 11\,095 points of the technique (ii) light curve on the
X-ray period given in table~1 results in the pulse profile shown by
the dotted line in the top panel of figure~\ref{fig2}. Note that the
phasing of both the technique (i) and (ii) profiles in
figure~\ref{fig2} can be directly compared, as all data have been
folded using the zero point given in table~1. As one would expect, the
technique (ii) light curve shown is in excellent agreement in terms of
morphology, phasing and pulsed fraction ($h_{rms}=26\pm8$\%) with the
technique (i) light curve, lending additional confidence to our
reduction and analysis techniques.

\section{Discussion and conclusions}
\label{disc}

The results presented in section~\ref{results} demonstrate
conclusively that we have detected optical pulsations from
1E\,1048.1--5937 on the X-ray spin period.

It is instructive to compare the optical light curve of
1E\,1048.1--5937 with that of the only other AXP to have been studied
in this way -- 4U\,0142+61 \citep{dhillon05}. Both objects show
optical pulsations on the X-ray spin period, which is 6.458\,s in the
case of 1E\,1048.1--5937 and 8.688\,s in 4U\,0142+61. Both objects
show optical pulsations with similar morphologies to their 2--10 keV
X-ray light curves, with 1E\,1048.1--5937 exhibiting a single-humped
pulsation and 4U\,0142+61 a double-humped pulsation.  Both objects
exhibit optical pulsations which are approximately in phase with the
X-ray pulsations, with 1E\,1048.1--5937 showing only marginal evidence
for the optical leading the X-rays and 4U\,0142+61 showing only
marginal evidence for the optical lagging the X-rays. Even the optical
pulsed fractions of the two objects are similar, with values of
$h_{pt}=52\pm15$\% and $h_{rms} = 21\pm7$\% in 1E\,1048.1--5937, and
$h_{pt}=58\pm16$\% and $h_{rms} = 29\pm8$\% in 4U\,0142+61.

The only major difference when comparing this study of
1E\,1048.1--5937 with Dhillon et al.'s (2005) study of 4U\,0142+61 is
the ratio of the optical to X-ray pulsed fraction: in 1E\,1048.1--5937
it is approximately unity, whereas in 4U\,0142+61 the optical
pulsations had an rms pulsed fraction 5--7 times that of the
X-rays. However, whereas the optical and X-ray pulsed fractions for
1E\,1048.1--5937 were measured contemporaneously, those of 4U\,0142+61
were not. The X-ray pulsed fractions of 4U\,0142+61 reported by
\cite{dhillon05} were quoted from the work of \cite{patel03}, who
obtained their {\em Chandra} data in 2000, over two years prior to the
optical observations. We now know from the work of \cite{dib07},
however, that 4U\,0142+61 exhibited an increase in its pulsed light of
36\% in the 2--10 keV band between 2002 and 2004. Hence it is possible
that at least part of the discrepancy between the optical/X-ray pulsed
fraction ratio in 1E\,1048.1--5937 and 4U\,0142+61 is due to
variability in the latter source. This hypothesis is further supported
by the fact that \cite{dhillon05} measured a value of
$h_{pt}=58\pm16$\% in September 2002 whereas \cite{kern02b} measured
$h_{pt}=27\pm8$\% in November 2001, although there are other possible
reasons for this discrepancy (see \citealt{dhillon05} for a
discussion). It is also possible that the discrepancy between the
optical/X-ray pulsed fraction ratio in 1E\,1048.1--5937 and
4U\,0142+61 is due to variability in 1E\,1048.1--5937: as discussed in
section~\ref{tech2ii}, the X-ray pulsed fraction of 1E\,1048.1--5937
is inversely proportional to the total 2--10 keV flux.  The optical
pulsed fraction might not vary in the same way, implying that the
optical/X-ray pulsed fraction ratio in 1E\,1048.1--5937 could be
variable and we just happened to observe it when it was unity.

Viewed in isolation, the data on 1E\,1048.1--5937 presented in this
paper do not allow us to discriminate between the magnetar and
fallback disc models, as the optical and X-ray pulsed fractions are
approximately equal, although the tentative evidence we present for
the optical pulses leading the X-rays is irreconcilable with
reprocessing. Arguably the most important result of this paper,
however, is that it confirms the existence of pulsed optical light in
a second AXP, 1E\,1048.1--5937, thereby demonstrating that the
properties of the optical emission in 4U\,0142+61 are not unique. The
observed similarities between the optical and X-ray emission in both
1E\,1048.1--5937 and 4U\,0142+61 indicate that closely related
populations of particles, located in the same region of the
magnetosphere, are probably responsible for the emission. This paper
has also highlighted the way forward for time-resolved optical studies
of these incredibly faint objects, as the data presented in this paper
would have been unobtainable had 1E\,1048.1--5937 been in a faint
state.  By targetting observations during bright X-ray states, it
should be possible to study other AXPs, and possibly also soft gamma
repeaters, in the optical part of the spectrum, although this will
still require access to sensitive, high-speed cameras like ULTRACAM on
the world's largest telescopes.

\section*{Acknowledgments}

ULTRACAM is supported by STFC grant PP/D002370/1. SPL acknowledges the
support of an RCUK Fellowship and STFC grant PP/E001777/1. TRM and CC
are supported under STFC grant ST/F002599/1.  VMK holds the Lorne
Trottier Chair in Astrophysics and Cosmology, a Canada Research Chair
and acknowledges support from an NSERC Discovery Grant, CIFAR and
FQRNT. RPM acknowledges STFC for support through its rolling grant
programme.  Based on observations collected at ESO, Chile.

\bibliographystyle{mn2e}
\bibliography{abbrev,refs}

\begin{thebibliography}{}

\bibitem[\protect\citeauthoryear{{Alpar}}{{Alpar}}{2001}]{alpar01}
{Alpar} M.~A.,  2001, ApJ, 554, 1245

\bibitem[\protect\citeauthoryear{{Beloborodov} \& {Thompson}}{{Beloborodov} \&
  {Thompson}}{2007}]{beloborodov07}
{Beloborodov} A.~M.,  {Thompson} C.,  2007, ApJ, 657, 967

\bibitem[\protect\citeauthoryear{{Chatterjee}, {Hernquist} \&
  {Narayan}}{{Chatterjee} et~al.}{2000}]{chatterjee00}
{Chatterjee} P.,  {Hernquist} L.,    {Narayan} R.,  2000, ApJ, 534, 373

\bibitem[\protect\citeauthoryear{{Dhillon}, {Marsh}, {Hulleman}, {van
  Kerkwijk}, {Shearer}, {Littlefair}, {Gavriil} \& {Kaspi}}{{Dhillon}
  et~al.}{2005}]{dhillon05}
{Dhillon} V.~S.,  {Marsh} T.~R.,  {Hulleman} F.,  {van Kerkwijk} M.~H.,
  {Shearer} A.,  {Littlefair} S.~P.,  {Gavriil} F.~P.,    {Kaspi} V.~M.,  2005,
  MNRAS, 363, 609

\bibitem[\protect\citeauthoryear{{Dhillon}, {Marsh}, {Stevenson}, {Atkinson},
  {Kerry}, {Peacocke}, {Vick}, {Beard}, {Ives}, {Lunney}, {McLay}, {Tierney},
  {Kelly}, {Littlefair}, {Nicholson}, {Pashley}, {Harlaftis} \&
  {O'Brien}}{{Dhillon} et~al.}{2007}]{dhillon07b}
{Dhillon} V.~S.,  {Marsh} T.~R.,  {Stevenson} M.~J.,  {Atkinson} D.~C.,
  {Kerry} P.,  {Peacocke} P.~T.,  {Vick} A.~J.~A.,  {Beard} S.~M.,  {Ives}
  D.~J.,  {Lunney} D.~W.,  {McLay} S.~A.,  {Tierney} C.~J.,  {Kelly} J.,
  {Littlefair} S.~P.,  {Nicholson} R.,  {Pashley} R.,  {Harlaftis} E.~T.,
  {O'Brien} K.,  2007, MNRAS, 378, 825

\bibitem[\protect\citeauthoryear{{Dib}, {Kaspi} \& {Gavriil}}{{Dib}
  et~al.}{2007}]{dib07}
{Dib} R.,  {Kaspi} V.~M.,    {Gavriil} F.~P.,  2007, ApJ, 666, 1152

\bibitem[\protect\citeauthoryear{Dib~et al.}{Dib~et al.}{2008}]{dib08}
Dib~et al. R.,  2008, in preparation

\bibitem[\protect\citeauthoryear{{Durant} \& {van Kerkwijk}}{{Durant} \& {van
  Kerkwijk}}{2005}]{durant05}
{Durant} M.,  {van Kerkwijk} M.~H.,  2005, ApJ, 627, 376

\bibitem[\protect\citeauthoryear{{Durant} \& {van Kerkwijk}}{{Durant} \& {van
  Kerkwijk}}{2006}]{durant06a}
{Durant} M.,  {van Kerkwijk} M.~H.,  2006, ApJ, 650, 1082

\bibitem[\protect\citeauthoryear{{Ertan}, {Erkut}, {Ek{\c s}i} \&
  {Alpar}}{{Ertan} et~al.}{2007}]{ertan07}
{Ertan} {\"U}.,  {Erkut} M.~H.,  {Ek{\c s}i} K.~Y.,    {Alpar} M.~A.,  2007,
  ApJ, 657, 441

\bibitem[\protect\citeauthoryear{{Gavriil} \& {Kaspi}}{{Gavriil} \&
  {Kaspi}}{2004}]{gavriil04}
{Gavriil} F.~P.,  {Kaspi} V.~M.,  2004, ApJ, 609, L67

\bibitem[\protect\citeauthoryear{Hulleman, van Kerkwijk \& Kulkarni}{Hulleman
  et~al.}{2000}]{hulleman00}
Hulleman F.,  van Kerkwijk M.~H.,    Kulkarni S.~R.,  2000, Nat, 408, 689

\bibitem[\protect\citeauthoryear{Hulleman, van Kerkwijk \& Kulkarni}{Hulleman
  et~al.}{2004}]{hulleman04}
Hulleman F.,  van Kerkwijk M.~H.,    Kulkarni S.~R.,  2004, A\&A, 416, 1037

\bibitem[\protect\citeauthoryear{{Kaspi}, {Gavriil}, {Chakrabarty}, {Lackey} \&
  {Muno}}{{Kaspi} et~al.}{2001}]{kaspi01}
{Kaspi} V.~M.,  {Gavriil} F.~P.,  {Chakrabarty} D.,  {Lackey} J.~R.,    {Muno}
  M.~P.,  2001, ApJ, 558, 253

\bibitem[\protect\citeauthoryear{Kern \& Martin}{Kern \&
  Martin}{2002}]{kern02b}
Kern B.,  Martin C.,  2002, Nat, 417, 527

\bibitem[\protect\citeauthoryear{{Mereghetti}}{{Mereghetti}}{2008}]{mereghetti%
08}
{Mereghetti} S.,  2008, A\&AR, 15, 287

\bibitem[\protect\citeauthoryear{Naylor}{Naylor}{1998}]{naylor98}
Naylor T.,  1998, MNRAS, 296, 339

\bibitem[\protect\citeauthoryear{Patel, Kouveliotou, Woods, Tennant, Weisskopf,
  Finger, Wilson, G{\"o}{\u g}{\"u}{\c s}, van~der Klis \& Belloni}{Patel
  et~al.}{2003}]{patel03}
Patel S.~K.,  Kouveliotou C.,  Woods P.~M.,  Tennant A.~F.,  Weisskopf M.~C.,
  Finger M.~H.,  Wilson C.~A.,  G{\"o}{\u g}{\"u}{\c s} E.,  van~der Klis M.,
   Belloni T.,  2003, ApJ, 587, 367

\bibitem[\protect\citeauthoryear{{Perna}, {Hernquist} \& {Narayan}}{{Perna}
  et~al.}{2000}]{perna00}
{Perna} R.,  {Hernquist} L.,    {Narayan} R.,  2000, ApJ, 541, 344

\bibitem[\protect\citeauthoryear{Press \& Rybicki}{Press \&
  Rybicki}{1989}]{press89}
Press W.~H.,  Rybicki G.~B.,  1989, ApJ, 338, 277

\bibitem[\protect\citeauthoryear{Smith, Tucker, Kent, Richmond, Fukugita,
  Ichikawa, Ichikawa, Jorgensen, Uomoto, Gunn, Hamabe, Watanabe, Tolea, Henden,
  Annis, Pier, McKay, Brinkmann, Chen, Holtzman, Shimasaku \& York}{Smith
  et~al.}{2002}]{smith02}
Smith J.~A.,  Tucker D.~L.,  Kent S.,  Richmond M.~W.,  Fukugita M.,  Ichikawa
  T.,  Ichikawa S.,  Jorgensen A.~M.,  Uomoto A.,  Gunn J.~E.,  Hamabe M.,
  Watanabe M.,  Tolea A.,  Henden A.,  Annis J.,  Pier J.~R.,  McKay T.~A.,
  Brinkmann J.,  Chen B.,  Holtzman J.,  Shimasaku K.,    York D.~G.,  2002,
  AJ, 123, 2121

\bibitem[\protect\citeauthoryear{{Tam}, {Gavriil}, {Dib}, {Kaspi}, {Woods} \&
  {Bassa}}{{Tam} et~al.}{2008}]{tam08}
{Tam} C.~R.,  {Gavriil} F.~P.,  {Dib} R.,  {Kaspi} V.~M.,  {Woods} P.~M.,
  {Bassa} C.,  2008, ApJ, 677, 514

\bibitem[\protect\citeauthoryear{{van Paradijs}, {Taam} \& {van den
  Heuvel}}{{van Paradijs} et~al.}{1995}]{vanparadijs95}
{van Paradijs} J.,  {Taam} R.~E.,    {van den Heuvel} E.~P.~J.,  1995, A\&A,
  299, L41

\bibitem[\protect\citeauthoryear{{Wang}, {Bassa}, {Kaspi}, {Bryant} \&
  {Morrell}}{{Wang} et~al.}{2008}]{wang08}
{Wang} Z.,  {Bassa} C.,  {Kaspi} V.~M.,  {Bryant} J.~J.,    {Morrell} N.,
  2008, ApJ, 679, 1443

\bibitem[\protect\citeauthoryear{Woods \& Thompson}{Woods \&
  Thompson}{2006}]{woods06}
Woods P.~M.,  Thompson C.,  2006, in Lewin W. H.~G.,  van~der Klis M.,  eds,
  Compact Stellar X-ray Sources. CUP, Cambridge, p.~547

\end{thebibliography}

\label{lastpage}

\end{document}